\documentclass[sigconf,table]{acmart}

\usepackage{amsmath,amsfonts}
\usepackage{algorithmic}
\usepackage{multicol}
\usepackage{graphicx}
\usepackage{xcolor}
\usepackage{textcomp}
\usepackage{enumitem}
\usepackage{soul}
\usepackage{booktabs,tabularx,enumitem,ragged2e}

\AtBeginDocument{%
  \providecommand\BibTeX{{%
    \normalfont B\kern-0.5em{\scshape i\kern-0.25em b}\kern-0.8em\TeX}}}

\copyrightyear{2022}
\acmYear{2022}
\setcopyright{acmcopyright}\acmConference[SACMAT '22]{Proceedings of the 27th ACM Symposium on Access Control Models and Technologies (SACMAT)}{June 8--10, 2022}{New York, NY, USA}
\acmBooktitle{Proceedings of the 27th ACM Symposium on Access Control Models and Technologies (SACMAT '22), June 8--10, 2022, New York, NY, USA}
\acmPrice{15.00}
\acmDOI{10.1145/3532105.3535017}
\acmISBN{978-1-4503-9357-7/22/06}

\begin{document}
\fancyhead{}
\title{BlueSky: Activity Control: A Vision for "Active" Security Models for Smart Collaborative Systems}

\author{Tanjila Mawla}
\affiliation{%
  \institution{Department of Computer Science, Tennessee Technological University} \city{Cookeville} \state{TN} \country{USA}
}
\email{tmawla42@tntech.edu}

\author{Maanak Gupta}
\affiliation{%
  \institution{Department of Computer Science, Tennessee Technological University}\city{Cookeville} \state{TN} \country{USA}
}
\email{mgupta@tntech.edu}

\author{Ravi Sandhu}
\affiliation{%
  \institution{Institute for Cyber Security (ICS) and NSF C-SPECC Center, \\University of Texas at San Antonio}\city{San Antonio} \state{TX} \country{USA}
}
\email{ravi.sandhu@utsa.edu}

\renewcommand{\shortauthors}{Mawla et al.}

\begin{abstract}
  Cyber physical ecosystem connects different intelligent devices over heterogeneous networks. Various operations are performed on smart objects to ensure efficiency and to support automation in smart environments. An Activity (defined by Gupta and Sandhu \cite{ACAC}) reflects the current state of an object, which changes in response to requested operations. Due to multiple running activities on different objects, it is critical to secure collaborative systems considering run-time decisions impacted due to related activities (and other parameters) supporting active enforcement of access control decision.  Recently, Gupta and Sandhu \cite{ACAC} proposed  Activity-Centric Access Control (ACAC) and discussed the notion of \textit{activity} as a prime abstraction for access control in collaborative systems. The model provides an \textbf{active} security approach that considers activity decision factors such as authorizations, obligations, conditions, and dependencies among related device activities. This paper takes a step forward and presents the core components of an ACAC model and compares with other security models differentiating novel properties of ACAC. We highlight how existing models do not (or in limited scope) support `active' decision and enforcement of authorization in collaborative systems. We propose a hierarchical structure for a family of ACAC models by gradually adding the properties related to notion of \textit{activity} and discuss states of an activity. We highlight the convergence of ACAC with Zero Trust tenets to reflect how ACAC supports necessary security posture of distributed and connected smart ecosystems. This paper aims to gain a better understanding of ACAC in collaborative systems supporting novel abstractions, properties and requirements.

\end{abstract}



\maketitle

\section{Introduction}
Cyber physical system (CPS) refers to a combination of computational and physical embedded devices like sensors, actuators, and cloud or edge computing capabilities where the devices participate collaboratively in performing specific tasks. In the era of technology and AI, cyber physical systems are becoming popular as human interactions with smart devices using cloud and edge computing are making life easier in various sectors. 
Farming, manufacturing, healthcare, vehicles, transportation, homes, are some of the domains, which can been impacted with this data and IoT driven futuristic smart solutions, offering convenience, and comfort to the end users. However, with growing connectivity, multi-user and device access along with AI-enabled operations, there are increasing cyber security risks which must be curtailed in depth, and mitigated as the systems get more exposed. Deployment of secure mechanisms for critical infrastructures and CPS is important to assure resilience and protection from adversaries including state supported entities or script kiddies.

In smart collaborative domains, the access rights to devices, applications and systems need to be authorized and require robust, fine grained and ``active" security models to limit unauthorized access and operations, ensuring the confidentiality, integrity and availability of these connected systems. By \textbf{active} security models, we mean models that approach security modeling and enforcement from the perspective (and abstraction) of activities or tasks in the collaborative systems, and consider contextual constraints and changing system dynamics to allow or deny requested operations.



In our vision, the ACAC \textbf{active} security model, will approach security modeling and enforcement from the perspective of connected activities, and as such, provide the abstractions and mechanisms for the active run-time management of security as ``activities'' progress to fulfill an automated collaborative task. An activity is a unit of work which is performed by a device, reflecting its current state. \textbf{Intuitively an activity is a long continuous event occurring for some time duration due to a requested operation.} It embodies the coarse-grained state of the entity, relevant to making decisions about authorized operations that transition amongst different activities of the entity. 


\subsection{Motivation for Activity Control Models in Connected Systems}
 Smart ecosystems are vulnerable to attackers since numerous devices, intelligent behaviors of those devices and sensor-based mechanisms are connected with the users or subjects through cloud or edge assisted network. We believe it
is necessary to adopt a different viewpoint to  develop access control approaches for such task and activity driven connected systems which work collaboratively to fully automate the entire ecosystem. 
 For example, smart heating devices such as thermostats like Nest or Ecobee, use sensors data to automatically adjust the routine temperature. They monitor the user’s real time location and turn the heating on and off. Video doorbell is a popular IoT smart home device which notifies user to receive video calls from their doorbells when someone is at the door. These can allow users to unlock the door remotely using smartphone apps. In these two examples, different activities involved are \textit{turning heating on and off}, \textit{video call from doorbell}, and \textit {unlocking the door}. To make the home more secured with IoT devices, condition of making video calls when someone is loitering near the home can be defined. Wireless IoT applications are also utilized in livestock monitoring as part of smart agriculture. Sensors can collect data regarding the location, well-being, and health of their cattle. To prevent the spread of diseases, sick animals are not allowed to stay with the herd. This is done using the help of IoT devices and the labor costs of the ranchers can be reduced as they can monitor the locations remotely and can take actions accordingly. Inter-dependency between activities on different devices can exist as added security feature. For example, \textit{injecting nitrogen fertilizer} at precise depths and intervals by automated tillers should be done before \textit{the drone, acting as seeder, places corn seeds directly in the fertilized soil}. Drone will be denied the access (for any operation) to corn field to place corn seeds if the soil is not recently fertilized. In these examples for smart IoT-based systems, the initiation or denial of \textit{activities} is dependent on other activities and conditions. 

 
 The recently proposed activity-centric approach for smart ecosystem by Gupta and Sandhu \cite{ACAC} motivated the idea of developing ACAC, discussed activity primitives, characterized activity relations, and defined a preliminary activity control expression notation. In this paper, we extend this proposed approach by elaborating the distinction of ACAC from other related models such as TBAC \cite{TBAC}, UCON \cite{UCONConf}, ACON \cite{ACON} and ABAC \cite{ABAC} that gives the readers a clear justification for why we need ACAC active security model for collaborative systems. We also propose model components for ACAC, a family of ACAC models with novel and converging (to related models) characteristics, incorporate states of an activity and show how ACAC is capturing the tenets of \textbf{Zero Trust} \cite{rose2020zero}, which has not been discussed in the afore-mentioned earlier work.

The main contributions of the paper towards the vision of creating active security model are as follows: 
\begin{itemize}[leftmargin=*]
  \item We highlight the key distinguishing factors with other related but fundamentally distinct models. These models were proposed with single enterprise-single administrative control into consideration, however, will not fit to collaborative and distributed systems with multiple devices, and subjects interacting and operating. In addition, the pool of objects is not fixed, and it is tedious task for any administrator to manage these distributed smart objects.
   \item We present our preliminary thoughts on various model components and states of activities that motivate the necessity of \textbf{activity-centric} access control model. 
  \item We show the hierarchical structure of ACAC models, as a family of model, which can be consolidated to a fine grained security model where notion of activity along with obligations, conditions and relationships of activities are bound together.
  \item We discuss convergence of ACAC and Zero Trust tenets.
\end{itemize}

This research builds upon previous work by Gupta and Sandhu \cite{ACAC}, which focused on the activity control framework by identifying different activities and their relationships. This work is fundamentally different where we make explicit distinctions from previous models which is critical to motivating this work, as well as discuss model components and an eventual family of ACAC models. 

The rest of the paper is organized as follows. Section \ref{sec:related} discusses background on access control in smart CPS, and elaborates on proposed solutions including access control models. Section \ref{sec:motivation} explains how the existing related models, primarily Task based (TBAC), Usage Control (UCON), Activity Control (ACON) and most recent Attribute Based (ABAC) access control models, are fundamentally different than the proposed activity-centric models for connected ecosystems. Section \ref{sec:family} highlights the key model components and framework for the family of ACAC models. The section also discuss the states of an activity in a collaborative system, and how it can transition from one state to another based on different conditions. Section \ref{sec:conc} discusses some future work and concludes the paper.



%

\section{Background and Related Works}
\label{sec:related}
In this section, we discuss the general requirements in collaborative cyber physical systems which are critical to design the access control models for different domains. Also, we review traditional access control models that have been widely used in covering some of the security considerations in connected ecosystems. Our key focus is to show how the proposed \textbf{ACAC (Activity-Centric Access Control)} model differs with the functionality and abstractions of previous models and how it can contribute to fulfilling the important requirements in collaborative and connected systems.
\subsection{Access Control Properties for Smart Collaborative Systems}
Collaborative ecosystems are designed to create an automated or semi-automated cyber infrastructure connected with large number of users and devices. Domains like smart home \cite{acsmarthome}, smart farming \cite{smartFarming}, smart manufacturing, and smart cars \cite{sacmat} having IoT devices, using Artificial Intelligence (AI) and networking technologies, need to restrict the access of users to protect the system from intruders and unexpected behaviours of the entities (processes, smart objects, applications etc.) in the system. There are fundamental requirements for the cyber systems \cite{CPS,sacmat, extendedACON} which need proper attention while designing an access control model and enforcement architecture. 
Javier and Rubio \cite{CPS} discussed some specific requirements that are important to meet during designing a secure CPS. In addition, lately Park et al. \cite{extendedACON} proposed seven design principles, and their evolution with changing dynamics and multi-domain/administered connected ecosystems. 

\begin{table*}[!t]
\centering
\caption{Comparison Overview of Features Proposed in ACAC Model}
\vspace{-2mm}
\label{tab5:sum}
\rowcolors{1}{lightgray!40!}{}
\scalebox{0.8}{%
\begin{tabular}{*{30}{|p{1.67cm}|p{1.5cm}|p{1.6cm}|p{1.67cm}|p{1.67cm}|p{1.67cm}|p{1.7cm}|p{1.8cm}|p{1.67cm}|p{1.67cm}|p{1.67cm}}}
\hline
\textbf{Models} & \textbf{Notion of Activity}& \textbf{Multiple Object 
Activities}& \textbf{Activities Concurrency}& \textbf{Activity Precedence}& \textbf{Activities Dependency}& \textbf{Incompatible 
Activities}& \textbf{Conditional Constraints}& \textbf{Activities Mutability}& \textbf{Run-time Authorization}& \textbf{Obligations}\\
\hline
TBAC&Yes&No&Yes&No&No&No&No&No&Yes&No
\\
\rowcolor{lightgray!40!}
\hline
UCON&No&No&No&No&No&No&Yes&No&Yes&Yes
 \\
 \hline
ACON&Yes&No&No&No&No&No&Yes&No&No&No
 \\
 \hline
ABAC&No&No&No&No&No&No&Yes&No&No&No
 \\
  \hline
  \hline
\textbf{ACAC}&YES&YES&YES&YES&YES&YES&YES&YES&YES&YES
 \\
\hline

\end{tabular} \vspace{-3mm}}
\end{table*}

\textbf{Dynamicity} is one of the important aspects for modern cyber physical systems. The variation in technologies and integrated applications involved in a cyber systems can change the workflow processes that dynamically change certain parameters. The access control systems should be designed in a way that can adapt to on the fly changes and dynamic behaviours. 
Access control mechanism should be \textbf{Scalable} to adapt introducing new users, devices and fine-grained or complex security policies. Also, these systems need to have the knowledge of which devices are connected and what resources are available in a certain situation. 
In addition, \textbf{Flexible Administration} in selecting which attributes or other authorization parameters such as activities, relationships etc. are used for defining the access to the resources, significantly improves the trust relationships between the resource owner and other entities in the connected systems.
\textbf{Quality of Service} in terms of the small response time needed to make access decisions is required for real time connected network of devices.
In practice, it is difficult to find a system that will see no delay in decentralized architecture. But, local or edge based access control mechanisms and parallel architectures can be implemented to reduce the delay time to get secure access to the resources in the connected ecosystem. Kim et al. in \cite{acsmarthome} described \textit{three} necessary steps for access control focusing on the flexibility and new entry of the devices, state of the devices and user/subject management for operating or accessing the devices. Apart from limiting security issues, our aim for proposed ACAC is to reduce unexpected loses, avoid harmful environmental situations and utilize the smart system advancements as it saves time and increases cost-efficiency.
\vspace{-3mm}
\subsection{Access Control Models and Mechanisms}
Access control models are paramount in making a decision for subjects to access the resources in a system. Security administrators that implement the access control models specify the requirements, components, workflow, users, resources and other parameters of the system which are important and factored in security decision. 
\vspace{-2mm}
\subsubsection{\textbf{Classical Access Control Models and Beyond}}
Several research works \cite{AC,DAC,ABAC,MAC,RBAC, EGRBAC, ABACGupta, ACAC} focused on decision requirements, access control modeling, and policy language generation for enterprise or distributed systems including cloud, IoT and CPS. 
Traditional access control models, such as Discretionary Access Control (DAC), Mandatory Access Control (MAC) and Role-based Access Control (RBAC) \cite{AC, DAC, MAC, RBAC} are primarily designed and widely used in enterprise applications with a defined set of resources and are difficult to adapt (and scale) for dynamic IoT-based distributed and multi-domain administered environments.

An enterprise-oriented access control model, TBAC \cite{taskBasedAuthorization, TBAC} captures active security management in workflow-based applications. 
TBAC encompasses the just-in-time permissions by executors in the process for the completion of a task while traditional access control model only encounters subject-object relationships as authorization rights without considering the larger operational context. The applicability of TBAC model is realistic for the enterprise applications. However, TBAC is not a reliable model when it comes to different kind of tasks/multiple objects or devices (such as IoT-based CPS) initiated by various sources like IoT devices, sensors, events etc. in  smart systems.  
Usage Control (UCON) \cite{UCONConf, UCONJourn} proposed by Sandhu and Park covers the traditional access control models DAC, MAC and RBAC while adding obligations and conditions as access parameters to allow or deny operations by a subject on an object. 
It considers mutability of attributes but does not capture the notion of activity and mutability of activities (as proposed in ACAC) that may run for a longer period of time. Later, a Risk Adaptive Access Control (RAdAC) model \cite{kandala2011attribute} analyzed the risk factors in decision making and showed the added security measures in UCON model. 
Jin et al. in \cite{ABAC} addressed proposed formal Attribute-Based Access Control (ABAC) \cite{hu2015attribute} model, considering the subject and object attributes as parameters along with security policies to make a decision. 
Inheritance of attributes and their values from parent groups is discussed in  \cite{gupta2016mathrm, reachability}. However, only considering attributes of entities as decision parameter does not yield as the desired active security enforcement. In Section \ref{sec:motivation}, we elaborate how these models fail to capture the requirements (including activity relationships) of smart collaborative CPSs.

\vspace{-4mm}
\subsubsection{\textbf{Access Control Solutions for Smart Collaborative Ecosystems}}
Several solutions \cite{EGRBAC,schuster2018situational,capBac,gupta2020access,HABAC,thakare2020parbac,bhatt2021attribute,abac-cc,ACAC,FedCAC,extendedACON, cathey2021edge, colombo2021regulating} have been proposed to extend classical models, and even propose new security models for connected distributed ecosystems such as IoT and CPS.
A comprehensive model for smart home IoT that extends RBAC model, known as the extended generalized role-based access control (EGRBAC) is proposed by Ameer et al. \cite{EGRBAC}. In smart home IoT, for local access, users directly communicate with the IoT devices and for remote access, users use the cloud services to access IoT devices. To prevent unauthorized accesses to devices, user-to-device communication is protected via role-assignment in the proposed EGRBAC model. 
Based on ABAC, another model for home IoT \cite{HABAC} is proposed for user-to-device interaction. Unlike EGRBAC \cite{EGRBAC}, HABAC model \cite{HABAC} can prevent prohibited operation by a user on an object at the assignment time where EGRBAC can do it only in the enforcement time. 
An attribute-based access control for Industrial Internet of Vehicles (IIoV) is proposed by Gupta et al. in \cite{ABACGupta}. This work uses the components of ABAC along with the concept of groups and tested the model prototype for Amazon Web Services (AWS) IoT platforms. 
A situational access control \cite{schuster2018situational} for IoT is proposed where an environmental situation oracle (ESO) contains the details about how a situation is sensed, inferred or actuated. 
An attribute-based communication control is proposed in \cite{abac-cc} where data flow control based on message attributes is integrated with the generic ABAC model to restrict sending messages to users. In this work, the dynamicity is controlled from the perspective of users not from the entities performing different operations. Gupta et al. \cite{gupta2020access} proposed a formal access control model for Google Cloud Platform (GCP) IoT and highlighted some fine grained extensions.



Dynamic situations trigger multiple users to request for same resource at the same time defining separate roles in RBAC. To mitigate the problem of inefficiency and ineffectuality in Microsoft Azure IoT Cloud, a Priority-Attribute Based RBAC Model is proposed by Thakare et al. in \cite{thakare2020parbac}. The authors used Azure resource manager (ARM) token associated with the attributes such as priority level, time, etc. which correspond to specific type of roles. He et al. \cite{he2018rethinking} proposed a device-centric and capability and relationship-based access control for home IoT based on a survey based study of 425 users. In this work, authors reflect the user’s capability based on contextual factors to operate on or actuate a device. This work focused on the IoT capabilities (devices’ actions) rather than on per-device operations. In IoT infrastructures, device-to-device communication can be compromised by various attacks. A certificate based lightweight access control and key agreement scheme for IoT devices (LACKA-IoT) is proposed by Das et al. \cite{das2019provably} where the model claims to protect the communication against device impersonation and man-in-the middle attack. However, due to the lack of proof, the claim is not true and that is analyzed in \cite{chaudhry2020secure} by Chaudhury et al.  Their iLACKA-IoT scheme is successful against device impersonation attack, man-in-the middle attack, replay attack, malicious device deployment, device physical capture attack and ephemeral secrets leakage attack (ESLA). The authors also proposed a demand response management (DRMAS) scheme \cite{chaudhry2020securing} based on certificates to provide the security against device impersonation attack, man-in-the middle attack and few other attack in smart grid.
A federated capability-based access control (FedCAC) is proposed\cite{FedCAC} to mitigate the shortcomings of traditional access control models in IoT. 


An activity-centric access control mechanism for social computing system is proposed by Park et al. \cite{ACON}. ACON framework encounters the user privacy based on their preferences and activities (which are short-lived in social networks) that can restrict the activities of other users in the social platforms. 
Park et al. extended this framework \cite{extendedACON} to adapt dynamic Smart and Collaborative Computing Systems (SCSs). This extended ACON defines activities of usage, control, service and decision. As suggested, a device cannot perform any control activity as it does not have a system-independent mind and just provides services. The ACAC \cite{ACAC} first addresses the need for activity control for smart and collaborative systems where activities on single and multiple devices may restrict the initiation of a new activity or may work as triggering events for the continuity or revocation of permissions for other activities. Activities may run on same or different devices.

\vspace{-3mm}
\section{Distinction from Other Security Models}
\label{sec:motivation}

\begin{table*}[t]
\setlength{\tabcolsep}{2pt}
\renewcommand{\arraystretch}{0.2}
\centering

\caption{Comparison of TBAC \cite{TBAC} and ACAC Models}
\vspace{-2mm}
\rowcolors{1}{lightgray!20!}{}
\scalebox{0.8}{%
\begin{tabular}{*{15}{|p{2.3cm}|p{5.7cm}|p{6.5cm}|p{7.1cm}}}

\hline
\textbf{Criteria} & \textbf{TBAC} & \textbf{ACAC}& \textbf{Example Use Case} \\
\hline
\vspace{0.5mm} Notion of Activity& 
\vspace{0.5mm}
\begin{itemize}[noitemsep,nolistsep,leftmargin=*]
  \item A task is a unit of work which can be performed by subjects in the system.
  \item TBAC focuses on just-in-time authorization of permissions to complete the task.
  \item Focuses on "when" permissions are activated.
\end{itemize}
&
\vspace{0.5mm}
\begin{itemize}[noitemsep,nolistsep,leftmargin=*]
  \item Activity reflects the prolonged state of continuous event happening on a device in smart system.
  \item A collaborative system can have multiple activities at a moment. 
  \item Focuses on run-time activity dependencies.
\end{itemize}
&
\vspace{0.5mm} \textit{Thermal imaging} is activated on the aerial drone by autonomous tractor only after the drone is done with the pest \textit{spraying}. Or \textit{turning on water sprinkler} is allowed only after ploughing a farm. Activities (thermal imaging or water sprinkling) cannot be activated without checking the requisite.


\\
\hline
\vspace{0.5mm} Decision Parameters& 
\vspace{0.5mm}
\begin{itemize}[noitemsep,nolistsep,leftmargin=*]
  \item Type-based, usage and validity counts of the executioner of the operation.
  \item Permissions are activated to the subjects on the fly (during run-time), but are pre-defined.
  \item No notions of obligation or condition are supported.
\end{itemize}
&
\vspace{0.5mm}
\begin{itemize}[noitemsep,nolistsep,leftmargin=*]
  \item Activities (or state) of devices along with conditions, obligations and dependence of activities.
  \item Control is at the activity level, not per device or subject.
  \item Activities may or may not be allowed at a point of time, offering multi-layer security check.
\end{itemize}
&
\vspace{0.5mm} In IoT-based smart waste management, sensor notifies the garbage lifting truck to clear the bin when it detects the overflow weight for a bin. Consequently, the truck starts clearing the bin. There is a condition and an order of activities (\textit{weighing} and \textit{lifting}) as decision parameters. Clearing the bin is based on a specific situation on sensing. Individual permission is not associated with an authorization step with a life-cycle or validity counts compared to TBAC.\\


\hline
\vspace{0.5mm} Operations on single or multiple objects by subjects&
\vspace{0.5mm} \begin{itemize}[noitemsep,nolistsep,leftmargin=*]
  \item Determines authorization  on a single object at a time.
  \item No activities on other objects are assessed for the authorization.
  \item Activation of permissions for subjects is per object in the workflow.
\end{itemize}
&
\vspace{0.5mm} \begin{itemize}[noitemsep,nolistsep,leftmargin=*]
  \item Multiple objects, maintaining sequence or concurrency or handling urgency of invocation or revocation, can be operated.
  \item To permit set of activities on multiple objects due to an event is possible if condition is satisfied.
\end{itemize}
&
\vspace{0.5mm} In smart home, when smoke spreads, the fire alarm rings and the emergency exits get automatically opened. Two permissions are executed on two separate objects for safety purpose. TBAC's abstractions of invoking permissions cannot explain the conditionally related activities on separate devices.

\\
\hline
\vspace{0.5mm} Relationships between activities in collaborative ecosystem & 
\vspace{0.5mm} \begin{itemize}[noitemsep,nolistsep,leftmargin=*]
  \item Supports existential, temporal and concurrency relations between authorization steps (aka events).
  \item These relationship does not control activities in the system.
\end{itemize}
&
\vspace{0.5mm} \begin{itemize}[noitemsep,nolistsep,leftmargin=*]
  \item Supports relationships such as temporary or emergency, precedence, conditional and incompatibility determine activity permissions.
\end{itemize}
&
\vspace{0.5mm} A bulldozer \textit{digging} is paused when gas sensor alarm siren is activated. TBAC model does not support this \textit{precedence} relation of one activity over another while ACAC encounters this relation.

 \\
 \hline
\vspace{0.5mm} Continuity of Activities & 
\vspace{0.5mm} \begin{itemize}[noitemsep,nolistsep,leftmargin=*]
  \item Continuity of any permission needs to check the life-cycle or usage count.
  \item Access is only possible until the validity count reaches the limit.
\end{itemize}
&
\vspace{0.5mm} \begin{itemize}[noitemsep,nolistsep,leftmargin=*]
  \item Contextual factors, attributes of subjects and objects, states of the devices define the continuity of activities which is more applicable in large scale IoT environments.
\end{itemize}
&
\vspace{0.5mm} A water sprinkler may only be allowed for 30 minutes in an entire day, and can only be turned on twice. This continuity of sprinkler activity will depend on when the sprinkler was activated. TBAC does not support this continuity.


\\
 \hline
\vspace{0.5mm} Mutability of Parameters & 
\vspace{0.5mm} \begin{itemize}[noitemsep,nolistsep,leftmargin=*]
  \item TBAC does not describe mutability either for attributes or activity.
  \item It only updates the usage count for permissions of the subjects, not the activity.
\end{itemize}
&
\vspace{0.5mm} \begin{itemize}[noitemsep,nolistsep,leftmargin=*]
  \item ACAC proposes mutability which changes the attributes for entities and activities before, on-execution or after execution of an activity.
  \item Usage limit is addressed for a particular activity which results in mutability by changing states of devices.
\end{itemize}
&
\vspace{0.5mm} In smart farming, the moisture level of soil changes when the water is sprayed on the crop field. \textit{Soil moisture level} is an attribute which is changed after \textit{water spraying} activity and this activity will stop when the moisture level is adjusted. This mutability is not discussed in TBAC but in ACAC.

\\
 \hline
\vspace{0.5mm} Framework Level and Application Domain & 
\vspace{0.5mm} \begin{itemize}[noitemsep,nolistsep,leftmargin=*]
  \item TBAC is an abstract level framework and built for enterprise applications like agent and workflow-based systems.
\end{itemize}
&
\vspace{0.5mm} \begin{itemize}[noitemsep,nolistsep,leftmargin=*]
  \item ACAC is designed for IoT-based smart and collaborative ecosystems with multiple objects. It focuses on activities as the prime notion to limit operations in the connected environments.
\end{itemize}
&
\vspace{0.5mm} Applications like sales-order processing, transactions in banking are enterprises for TBAC. On the other hand, smart farming, smart home, smart healthcare monitoring, IoT based manufacturing factory etc. are the applications for ACAC. 

 \\
\hline

\end{tabular}

}
\label{tab1:tbac-acac}
\end{table*}
Our proposed Activity-centric Access Control focuses on notion of activity. However, in addition to assessing the activities performed on devices and dependencies among them, the model is convergent and closely related to few other related models TBAC \cite{TBAC}, UCON \cite{UCONConf}, ACON \cite{ACON} and ABAC \cite{ABAC}. Although, these prior models have different decision parameters and were primarily designed for single administration enterprises, it is critical to make explicit distinctions with these models to comprehensively motivate the need for ACAC. In this section, we review the related models, different distinguishing features and the limitations of existing models, which is critical to justify the need of proposed ACAC. 

Broadly, Table \ref{tab5:sum} summarizes the distinguishing features among ACAC and other compared models to understand the key motivation behind our proposed active security model for collaborative smart ecosystems, and how the existing models do not fulfill the requirements. This table shows how ACAC is significantly different yet convergent to other related models. 
The columns in the table refer to considering the following factors in the models mentioned in the first column: Notion of activity (which is different in other models as discussed later), consideration of activities on multiple objects while taking access decision, relationships between activities: concurrency, dependency, incompatibility, conditions, Mutability of activities (changing states of one or multiple activities on same or different devices to accept the request of another activity), presence of run-time authorization and obligations (required actions before or during execution-time of an event). 

Table \ref{tab1:tbac-acac}, \ref{tab2:ucon-acac}, \ref{tab1:acon-acac} and \ref{tab1:abac-acac} respectively highlights the major differences of TBAC \cite{TBAC}, UCON \cite{UCONConf}, ACON \cite{ACON} and and ABAC \cite{ABAC} with ACAC. The first column describes the differentiating criteria between other models and ACAC, the later two columns elaborate each criteria for other models and ACAC respectively and the final column explains the distinctions by an use case.

\begin{table*}[t]
\setlength{\tabcolsep}{2pt}
\renewcommand{\arraystretch}{0.2}
\centering

\caption{Comparison of UCON \cite{UCONJourn} and ACAC Models}
\vspace{-2mm}
\rowcolors{1}{lightgray!20!}{}
\scalebox{0.8}{
\begin{tabular}{*{15}{|p{2.5cm}|p{5.3cm}|p{6.7cm}|p{7cm}}}
\hline
\textbf{Criteria} & \textbf{UCON} & \textbf{ACAC}& \textbf{Example Use Case} \\
\hline
\vspace{0.5mm} Notion of Activity &
\vspace{0.5mm}
\begin{itemize}[noitemsep,nolistsep,leftmargin=*]
  \item No notion of activity.
  \item Designed for digital rights management.
  \item UCON discuss the access to a single object along with usage control.
\end{itemize}
&
\vspace{0.5mm}
\begin{itemize}[noitemsep,nolistsep,leftmargin=*]
  \item Activities are the prolonged state of object entities due to an operation.
  \item Activities can be requested by a source like user, device, sensor or can be triggered by any event.
  \item Focus on smart collaborative and automated systems.
\end{itemize}
&
\vspace{0.5mm} When a nutrient solution is getting mixed into the crop field, pesticide spraying cannot be turned-on. Here, activity of mixing nutrient ingredient is making the activity of spraying pesticide to wait until the previous activity is finished. This notion of activity is not included in UCON, as activity is a decision factor.

 \\
\hline
\vspace{0.5mm} Decision Parameters &
\vspace{0.5mm}
\begin{itemize}[noitemsep,nolistsep,leftmargin=*]
  \item Subject and object attributes.
  \item Authorizations
  \item Obligations
  \item Conditions
\end{itemize}
&
\vspace{0.5mm}
\begin{itemize}[noitemsep,nolistsep,leftmargin=*]
\item Activity relationships such as concurrent and temporal relations between multiple objects/devices.
  \item Authorizations, obligations, conditions.
\end{itemize}
&
\vspace{0.5mm} When \textit{weed removal} is necessary in a crop field, it needs to be done before starting \textit{spraying} nutrient ingredients so that this dependency can help reducing the wastage of nutrient solution. Here, one activity is delaying another activity. This constraint is not considered in UCON.

 \\

\hline
\vspace{0.5mm} Consideration of Multiple objects &
\vspace{0.5mm}
\begin{itemize}[noitemsep,nolistsep,leftmargin=*]
  \item Permission is per single object.
  \item Decision of accessing one object is not impacted by operations or access rights of subjects on other objects.
\end{itemize}
&
\vspace{0.5mm}
\begin{itemize}[noitemsep,nolistsep,leftmargin=*]
\item State of multiple objects may need to be checked since state of the IoT devices can constrain activities in the system.
\end{itemize}
&
\vspace{0.5mm} Weed removal and pesticide spraying on same crop field by same or separate drones cannot occur concurrently as they are contradictory in terms of functionalities. UCON model does not limit operations based on state of different devices.

\\
\hline
\vspace{0.5mm} Dependencies based on the activity relationship (Temporal, Precedence, Dependence, incompatibility etc.) \cite{ACAC}.
&
\vspace{0.5mm}
\begin{itemize}[noitemsep,nolistsep,leftmargin=*]
\item Authorizations (A), obligations (B) and conditions (C) are the prime usage decision components. 
\item No dependencies among activities or events are considered.
\end{itemize}
&
\vspace{0.5mm}
\begin{itemize}[noitemsep,nolistsep,leftmargin=*]
\item To initiate, continue or stop an activity, status of other relevant activities are assessed which is marked as dependencies (D) in the model components.
\item ACAC involves all four parameters, and can be referred as $\mathrm{ACAC_{ABCD}}$.
\end{itemize}
&
\vspace{0.5mm} There may be sequential relation between the initiation of tasks on machine A and machine B in a smart system for a certain interval. Machine A needs to be started before machine B starts. UCON does not consider this dependency but ACAC recognizes this relation prior to the initiation of an activity.

 \\
 \hline
\vspace{0.5mm} Mutability of Parameters & 
\vspace{0.5mm}
\begin{itemize}[noitemsep,nolistsep,leftmargin=*]
\item Considers mutability of attributes.
\item Does not consider about changing the state of any running activity due to change in attributes.
\end{itemize}
&
\vspace{0.5mm}
\begin{itemize}[noitemsep,nolistsep,leftmargin=*]
\item ACAC defines mutability of attributes as well as mutability of activities, to allow an activity only for certain time or usage purpose.
\end{itemize}
&
\vspace{0.5mm} Playing song on Alexa is an activity and is only allowed to play for twice a day irrespective of which user starts this activity. This is mutability of activity capturing the usage count of activity.
 \\
\hline
\end{tabular}}
\label{tab2:ucon-acac}
\end{table*}

\vspace{-2mm}

\section{Foundations of Activity-Centric Model for Collaborative Ecosystem}
\label{sec:family}
 The proposed ACAC model 
 offers convergence and extends previously defined models (discussed in previous section) considering aspects like obligations, conditions, mutability of activities in addition to several new abstractions and features focusing on run-time and active security considerations in connected systems. In this section, we first describe the identified ACAC model components and discuss how incrementally we will get a family of models having different levels of flexibility and expressiveness in terms of access control policy definitions needed for ``active" run time security model for collaborative systems.
\begin{table*}[t]
\setlength{\tabcolsep}{1pt}
\renewcommand{\arraystretch}{0.2}
\centering
\caption{Comparison of ACON \cite{ACON} and ACAC Models}
\vspace{-2mm}
\rowcolors{1}{lightgray!20!}{}
\scalebox{0.8}{%
\begin{tabular}{*{15}{|p{2.3cm}|p{5.7cm}|p{6.5cm}|p{7.4cm}}}
\hline

\textbf{Criteria} & \textbf{ACON}& \textbf{ACAC}& \textbf{Example Use Case} \\
\hline
\vspace{0.5mm} Notion of Activity & 
\vspace{0.5mm} \begin{itemize}[noitemsep,nolistsep,leftmargin=*]
  \item Short-lived activities in the social computing context. Not reflecting state of object.
  \item Activity examples: \textit{poke} a user, \textit{tag} a friend.
\end{itemize}
&
\vspace{0.5mm}
\begin{itemize}[noitemsep,nolistsep,leftmargin=*]
  \item Long-lived activities are requested by user or any other entity in automated system.
  \item Activities are performed by the devices reflecting their state.
\end{itemize}
&
\vspace{0.5mm} \textit{Planting seeds} is an activity that is performed by drone in smart farming without user's intervention. ACON's target object do not change the state, controlled by users or systems.

 \\
\hline
\vspace{0.5mm} User Privacy Preference & 
\vspace{0.5mm}
\begin{itemize}[noitemsep,nolistsep,leftmargin=*]
  \item Privacy setting from user and administrator can control the activities on objects such a posts, comments etc. performed in the social system.
\end{itemize}
&
\vspace{0.5mm}
\begin{itemize}[noitemsep,nolistsep,leftmargin=*]
  \item Control is at the activity level based on the states of devices, not per user, administrator or resource.
\end{itemize}
&
\vspace{0.5mm} To turn the smart speaker on by a child needs to check whether parents are watching TV in the living room or not. Decision based on other activities is not in the scope of activity control in ACON model.

\\

\hline
\vspace{0.5mm} Authorization& 

\begin{itemize}[noitemsep,nolistsep,leftmargin=*]
  \item Activity is controlled depending on the pre-authorization defined for user or administrator.
  \item These are passive permissions, and do not consider the context of the request.
\end{itemize}
&
\begin{itemize}[noitemsep,nolistsep,leftmargin=*]
  \item Both pre-authorization and ongoing-authorization control the activities in the system. 
  \item Multi-level checks are supported, one based on policies and other on run-time dependencies.
\end{itemize}
&
\vspace{0.5mm} Smart floor cleaner updates number of objects on the floor. It starts \textit{cleaning} when it detects number of objects above threshold and stops \textit{cleaning} when number of objects reduces below threshold. This ongoing authorization process of performing \textit{cleaning} activity. Continuous update of the state of object and activity is not considered in ACON.
 \\
 
 \hline
\vspace{0.5mm} Relationship Among Activities & 
\begin{itemize}[noitemsep,nolistsep,leftmargin=*]
  \item No connection between activities to control new activities requested in the system.
\end{itemize}
&
\begin{itemize}[noitemsep,nolistsep,leftmargin=*]
  \item Dependence among activities is a critical decision factor considered to allow or deny a requested activity.
\end{itemize}
&
\vspace{0.5mm} Thermal imaging is activated on the Drone by autonomous tractor only after the drone is done with thespray, and as a consequence of the start of Thermal-imaging, the spraying should be stopped.

 \\

 \hline
\vspace{0.5mm} Mutability of Parameters &
\vspace{0.5mm}
\begin{itemize}[noitemsep,nolistsep,leftmargin=*]
  \item No policy or attributes change during the lifetime of an activity.
  \item Activities are not mutable.
\end{itemize}
& 
\begin{itemize}[noitemsep,nolistsep,leftmargin=*]
  \item Mutability of attributes allows update of the attributes as a side-effect of any event or activity
  \item Mutability of activities allows devices to start, abort, halt and hold any activity.
\end{itemize}
&
\vspace{0.5mm} The moisture level of soil is adjusted after the water spraying activity for a duration of time. Update of the attribute (moisture level) of an entity (soil) changed the activity state (stopping water spraying). The automation of change of device's state cannot be explained in ACON.

 \\
\hline
\end{tabular}
\vspace{-2mm}}
\label{tab1:acon-acac}
\end{table*}
\begin{table*}[t]
\setlength{\tabcolsep}{2pt}
\renewcommand{\arraystretch}{0.1}
\centering
\caption{Comparison for ABAC \cite{ABAC} and ACAC Models}
\vspace{-2mm}
\renewcommand{\arraystretch}{0.6}
\rowcolors{1}{lightgray!20!}{}
\scalebox{0.8}{%
\begin{tabular}{*{15}{|p{2.3cm}|p{5cm}|p{6.6cm}|p{7.6cm}}}
\hline

\textbf{Criteria} & \textbf{ABAC}& \textbf{ACAC}& \textbf{Example Use Case} \\
\hline

\vspace{0.5mm} Notion of activity & 
\begin{itemize}[noitemsep,nolistsep,leftmargin=*]
  \item No notion of activity
  \item Actions that are performed by subjects on objects need access decisions.
\end{itemize}
& 
\begin{itemize}[noitemsep,nolistsep,leftmargin=*]
  \item Activity in ACAC: active states of objects
  \item Dependence between activities determines if a requested activity is allowed
\end{itemize}
&
\vspace{0.5mm} 
When a nutrient solution is getting mixed into the crop field, pesticide spraying cannot be turned-on. Access limitation like this example due to an activity constraint is out of scope for ABAC.

 \\
\hline
\vspace{0.5mm} Decision Parameter &
\begin{itemize}[noitemsep,nolistsep,leftmargin=*]
  \item Attributes of the entities along with pre-defined security policies
  \item Environmental conditions
\end{itemize}
&
\begin{itemize}[noitemsep,nolistsep,leftmargin=*]
  \item Authorizations (could be pre-defined ABAC or other models)
  \item Obligations
  \item Conditions
  \item Dependence among activities (state of devices)
\end{itemize}
&
\vspace{0.5mm} Attribute such as the temperature of a facility can determine the initiation of air cooler, but the state of thermostat can limit the air-cooling. Thermostat needs to be turned on at the same time with air-cooling. Two related activities are not considered as decision factors in ABAC.

 \\

\hline
\vspace{0.5mm} Active Security Management & 
\begin{itemize}[noitemsep,nolistsep,leftmargin=*]
  \item Not an active security model: once the decision is determined based on policy, it cannot be changed during a session.
\end{itemize}
&
\begin{itemize}[noitemsep,nolistsep,leftmargin=*]
  \item Active security model: enforces continuous decision real time working environment based on attributes, conditions, obligations, activities and dependence among them.
\end{itemize}
&
\vspace{0.5mm} Updating the number of food-packet sealed using a robotic arm is required. The number during the packaging session decides the continuity of \textit{packaging}. This automatic decision process is an active task management in ACAC.

 \\
\hline
\vspace{0.5mm} Obligations as Decision Factors &
\begin{itemize}[noitemsep,nolistsep,leftmargin=*]
  \item No follow-up or pre- operation or event is required by the subject for an access request.
 \end{itemize}
&
\begin{itemize}[noitemsep,nolistsep,leftmargin=*]
  \item Obligations may be needed or performed on the same/different objects by the requesting source or subject.
 \end{itemize}
 &
 \vspace{0.5mm} Fingerprint verification is needed to confirm the age before ordering online through Alexa. This is an obligation needs to be performed by the accessing subject. Obligations are not supported in ABAC.

 \\

 \hline
\vspace{0.5mm} Mutability of Parameters &
\begin{itemize}[noitemsep,nolistsep,leftmargin=*]
  \item Attributes changed after execution of any authorized right is not considered.
\end{itemize}
&
\begin{itemize}[noitemsep,nolistsep,leftmargin=*]
  \item Mutability of attributes allows update of the attributes as a side-effect of any event or activity.
  \item Mutability of activities or their usage can limit current or future new activities.
\end{itemize}
&
\vspace{0.5mm} Pest spray for particular number of hours per week needs to check the updates of hour after spraying every time. This update limits the \textit{spraying activity} for the remaining hours in a week. ABAC does not discuss mutability but ACAC does.

 \\
\hline
\end{tabular}
}
\label{tab1:abac-acac}
\vspace{-4mm}
\end{table*}

\subsection{ACAC Model Components}\label{modelComponents}

\begin{figure}[t!]
    \centering
    \includegraphics[width=\columnwidth]{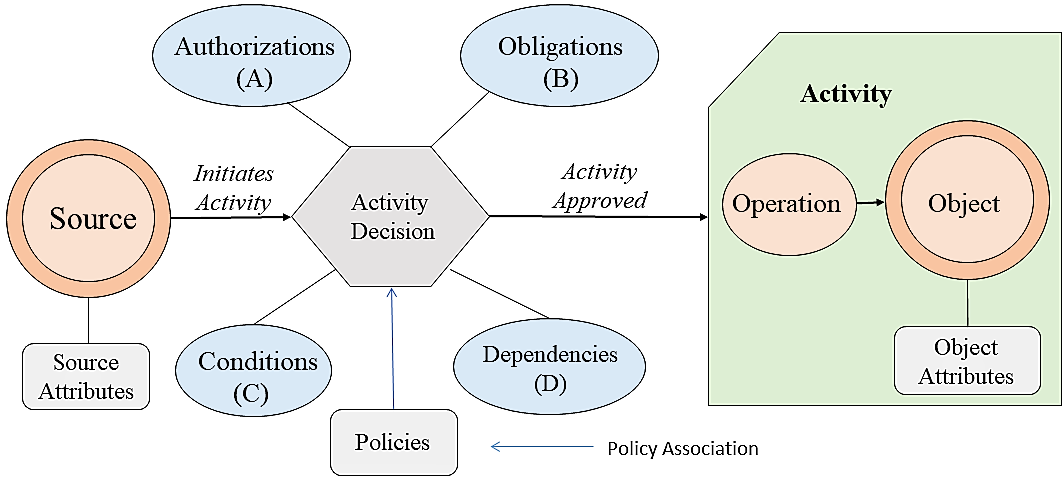}
    \centering
    \caption{ACAC Model Components.}
    \vspace{-5mm}
    \label{fig-model}
\end{figure}

ACAC consists of the following \textbf{components}: sources, source attributes, objects, object attributes, activities, authorizations (A), obligations (B), conditions (C), dependencies (D) based on relationship of activities (characterized by Gupta and Sandhu \cite{ACAC}), policies and operations, as shown in Figure \ref{fig-model}. 
In general, when a source initiates an activity, which will be started by an operation on the corresponding object, the activity decision is made considering four parameters (A-B-C-D) which are reflected in the blue eclipses. After the activity is allowed, the requested operation will be performed on the object entity which is separately shown in the green colored box shaped area.

\textbf{Source} is the initiator of an activity (explained below) on objects, devices and applications in the ecosystem. Source can be a device, an user, a sensor, or a process in the system. An event can also trigger an activity to be started on an object. For instance, in case of a oil leak event, different activities can be triggered such as wiping floor with cleaner, detecting oil level etc. \textbf{Object} is an entity on which the operation is performed and must be protected. Objects in smart ecosystems include devices, sensors, applications or other technologies that are connected in the IoT based system. 
In agriculture, objects include tractor, drone, livestock monitoring device or  camera. Objects are accessible directly by the source or it can be accessed remotely through an application.
\textbf{Operations} are the actions that change the state of an object entity. 
An \textbf{activity} on an object refers to the state (Intuitively an activity is a long continuous event occurring for some time) of the object that is changed due to an operation performed by various sources where the action is allowed based on conditions, obligations and dependencies among activities.  In the previous example, \textit{wiping floor} is the activity which is started by an operation initiated due to oil leak event (which could be detected by a sensor itself). Here, \textit{Turn-on} wiper is an operation that starts the activity \textit{wiping} on the floor.

\textbf{Source and object attributes} contribute in the decision process of ACAC. Attributes reflect the characteristic of the entities. Source and objects attributes can be mutable, which can be changed before, during or after an activity is invoked. In smart agriculture, the humidity of greenhouse needs to be adjusted using humidifier. A maximum threshold level of humidity allows the humidifier to be turned-off. Humidity is a mutable attribute in this example.

\textbf{Authorizations (A)} are based on the source and object attributes, and the right of performing the corresponding operation related to the requested activity on an object by a source. Authorizations are static as well as predefined and evaluated prior to the access. However, in a special case of revocation of access due to attribute mutability, authorizations can be ongoing. For instance, in smart industry, the filtering process on Oil Filter can be turned on by the production manager. Here, role of the user contributes as source attribute in the authorization process for turn-on operation. Authorization is usually required prior to the access, but in addition it is possible 
to require ongoing authorization during the access. 
\textbf{Obligations (B)} are the requirements that must be fulfilled by the source to access any object. For example, fingerprint enabled door lock needs the valid entry of user's fingerprint to access the house. Here, the obligation is scanning the finger and thus, user identity is verified. Obligations can be related activity or an action required to fulfill the requested activity.
\textbf{Conditions (C)} are the system or environmental factors related to the requesting operation. The conditions could be pre- or current conditions, which may also capture the usage count for particular activity. In agricultural industry, Subsurface Drip Irrigation (SDI) system can help the farmers to understand the exact time the plants need to be watered. Here, the condition can involve the temperature and soil moisture level as parameters to measure the time and amount of water to be sprayed on crops. \textbf{Dependencies (D)} based on relationship of activities that are running on single or multiple entities is a critical factor in the ACAC active security model. When a source request an activity to start, it may check the status of other activities to avoid conflict and loses or system failures. For example, in greenhouse automation, when the air-cooler is turned on, the humidifier must be turned on at the same time to make sure that there is a consistent humidity prevailing in the greenhouse. This concurrent relation needs to be fulfilled for starting the \textit{cooling} activity.

\textbf{Activity Decision} indicated in the Figure \ref{fig-model} determines when an activity is allowed in the system based on the attributes, authorization (A), obligations (B), conditions (C) and dependencies (D) based on relationship among activities in the collaborative system. In UCON model \cite{UCONJourn}, three decision factors (A, B, C) are considered (usually referred as $\mathrm{UCON_{ABC}}$), and do not consider the active run-time access control decision. By adding dependencies (D), ACAC model provides the means for \textbf{``active"} decision management, and hence, can also be referred as $\mathrm{ACAC_{ABCD}}$. By active decision management, we mean the model takes the control decision from the perspective of activities along with other related parameters. In active approach, activities on different objects are invoked, constantly monitored and revoked depending on the contextual factors of run-time environment. Activity decision function returns true or false based on different decision parameters as stated above. The concept of activity usage, along with active monitoring of our related activities together with conditions can be tracked in real time supporting active security model. This function is expressed as \textbf{activity control policies} using a policy language. This language can be developed using propositional logic capturing different components required to make an activity decision. An early attempt was made by Gupta and Sandhu \cite{ACAC} which can lead us to build formal policies for ACAC. We aim to formalize the policies for our proposed model which can be generalized for smart domains in our future work.

\subsection{A Family of ACAC Models}\label{AA}
The core idea of ACAC lays the foundation for a model that takes run-time access  control decision focusing on activity occurring in the system and their dependencies. Instead of designing a single monolithic model with different supporting abstractions, we have created a family of models that gradually adds the features and finally builds a consolidated model. 

\begin{figure}[!t]
    \centering
    \includegraphics[width=\columnwidth]{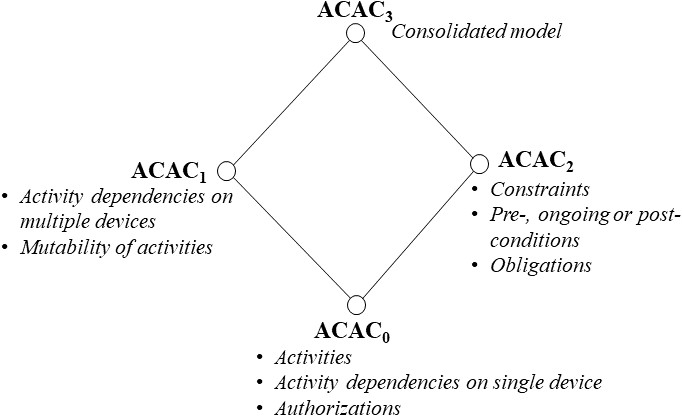}
    \centering
    \caption{A Framework for a Hierarchy of ACAC models}
    \vspace{-5mm}
    \label{fig:family}
\end{figure}

\textbf{Hierarchy of ACAC Models:}
Figure \ref{fig:family} shows the framework for a hierarchy of ACAC models. ACAC$_0$ is the base model which is shown at the bottom of the hierarchical structure. It incorporates the minimum requirements to develop activity-centric access control model supporting run-time authorization of activities along with dependencies (on same/single device) in the collaborative ecosystem. ACAC$_1$ and ACAC$_2$ inherit the general and flexible ACAC$_0$ model and add more explicit features. ACAC$_1$ encompasses activity dependencies on multiple devices, mutability of activities and device to device authorizations. ACAC$_2$ adds \textbf{two} decision parameters (Conditions and Obligations) which are required in access control model for smart ecosystems. ACAC$_2$ also supports constraints which are classified as static and dynamic constraints discussed later. The consolidated model ACAC$_3$ includes ACAC$_1$ and ACAC$_2$ which hierarchically gets the characteristics of ACAC$_0$. With this family of models, we show how ACAC is a novel model for smart collaborative systems and synergistically converging to other related models.

\begin{figure}[!t]
    \centering
    \includegraphics[width=\columnwidth]{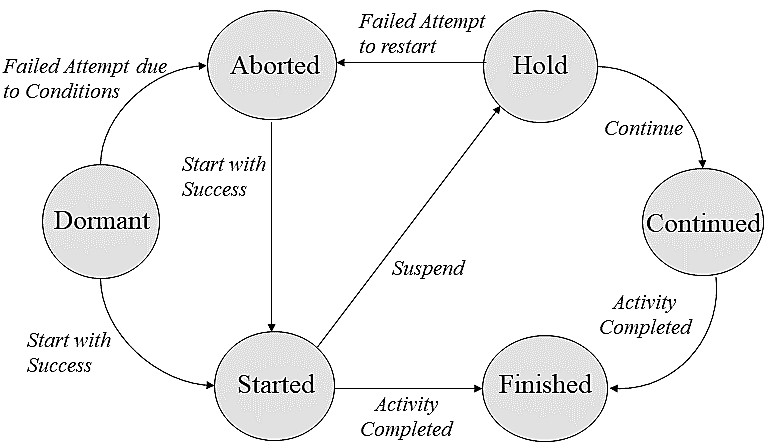}
    \caption{States of an Activity}
    \vspace{-4mm}
    \label{fig:activity-state}
\end{figure}

\textbf{The Model ACAC$_0$:}
We describe ACAC$_0$ here. We explain an activity along with states of an activity (i.e. life-cycle of an activity once it is requested and started, what different states it can transition due to different factors), user to device authorization and activity dependencies on a single device.
\textbf{States of an Activity and life-cycle:} Activity is the result of an operation that changes the state of a device. An activity can have multiple states during its lifetime due to an event, user operation, activity dependencies, conditions, obligations or other factors. Figure \ref{fig:activity-state} reflects the \textbf{states of an activity} showing the transition of an activity into different states during its life-cycle as described below.

\begin{itemize}[leftmargin=*]
  \item \textbf{Dormant:} Activity in this state means that the activity is not invoked yet. The dormant state is viewed as an inactive state while the activity is requested and the related dependencies are being assessed to see if the corresponding operation can be performed on the target device or not.
  \item \textbf{Started:} Activity is successfully invoked and target object is performing the function. During this state, the related attributes of the object and system can be changed and contribute to deciding the next state of the running object.
  \item \textbf{Aborted:} When a failed attempt is made to start an activity. This state is similar to the dormant state except bringing the requested activity to a premature end due to certain conditions including activity dependencies.
  \item \textbf{Hold:} When a running activity is temporarily suspended for some reason and it may be invoked again or continue after certain condition is met.
  \item \textbf{Continued:} When an activity is running after a temporary suspension.
  \item \textbf{Finished:} In this state, the activity is completed and the access to this activity on the performing device is revoked. 
\end{itemize}

\textbf{Authorizations:} To protect devices an dother objects from unauthorized access, our base model ACAC$_0$ provides the means to specify authorization policies for users and devices. Authorization is static as well predefined and it binds permissions (for operations) to specific users and devices based on their selected attributes. Authorizations can evaluate attributes of entities with a set of authorization rules
for activity decision. This can be pre or ongoing authorization.
\textbf{Activity Dependencies on Single Object:} When an activity is requested to be initiated on a device, the state of the device and dependencies with other activities on that same device must be evaluated. Gupta and Sandhu \cite{ACAC} discussed relationship and characterization of activities such as ordered, concurrent, or incompatible in more details.

\textbf{The Model ACAC$_1$ to support Mutability of Activities:}
\textbf{Mutability of activities} refers to changing the state of devices based on usage and the dependencies with other activities on same or multiple different devices. We specify usage count as one of the conditional factors that limits the count and duration of running an activity. For example, spraying nutrient solution on crops may be needed for 2 hours per week and spraying more than 30 minutes is not allowed per day. When an aerial drone is turned-on to spray nutrients, spraying time is continuously checked to make sure that the running time is not exceeding 30 minutes and the total duration in a week is less than 2 hours. Mutability will restrict the dependent and related activities.  
Table \ref{tab6:mutability} reflects how one or more activities (on single or multiple devices) are related and mutable with a requested activity in terms of the invocation time. We put the relationships (explained in \cite{ACAC}) in the first column. 
Immutable property in the second column denotes if activities involved in the mentioned relations have any dependency of changing their states or not. \textit{Pre-invocation} indicates that at least one activity may be required to be started before the invocation of a requested activity, \textit{parallel invocation} requires at least one other activity to be concurrently running with the requested activity and \textit{post-invocation} means that at least one activity is required to be started after the requested activity is finished. For example, \textbf{Order} relation between two activities can be mutable (marked as $\surd$ under pre- and post-invocation). A requested activity may need another activity(s) to be finished or started after it. \textbf{Concurrent} relationships refer to the execution of two or more activities including the requested one in parallel (marked as $\surd$ under parallel invocation). These dependencies are part of the mutability as states of several activities on multiple objects can be changed in order accommodate the requested one.

\begin{table}[!t]
\centering
\caption{Mutability of Dependent Activities in terms of the invocation time related to a requested activity. $\surd$ and $\times$ respectively denote the presence (mandatory or optional) and absence of the corresponding field to support the relationships in the first column.}
\vspace{-2mm}
\rowcolors{1}{lightgray!20!}{}
\scalebox{0.9}{%
\begin{tabular}{*{10}{p{1.4cm}}}
\hline

\textbf{Activities Relationship} & \textbf{Immutable}& \textbf{Pre-invocation}& \textbf{Parallel invocation} & \textbf{Post-invocation}\\
\hline
Independent & \;\;\;\;\; $\surd$ & \;\;\;\;\; $\surd$ & \;\;\;\;\; $\surd$ & \;\;\;\;\; $\surd$
 \\
\hline
Ordered  & \;\;\;\;\; $\times$ &\;\;\;\;\;  $\surd$ & \;\;\;\;\; $\times$ & \;\;\;\;\; $\surd$
\\
\hline
Concurrent & \;\;\;\;\; $\times$ &\;\;\;\;\;  $\times$ & \;\;\;\;\; $\surd$ &\;\;\;\;\;  $\times$
 \\
 \hline
Temporary & \;\;\;\;\; $\times$ & \;\;\;\;\; $\surd$ & \;\;\;\;\; $\surd$ & \;\;\;\;\; $\surd$
 \\
  \hline
Precedence & \;\;\;\;\; $\times$ & \;\;\;\;\; $\times$ & \;\;\;\;\; $\times$ & \;\;\;\;\; $\surd$
 \\
 \hline
Conditional & \;\;\;\;\; $\times$ &\;\;\;\;\;  $\surd$ &\;\;\;\;\;  $\surd$ &\;\;\;\;\;  $\surd$
 \\
 \hline
Incompatible & \;\;\;\;\; $\times$ & \;\;\;\;\; $\times$ &\;\;\;\;\;  $\times$ & \;\;\;\;\; $\times$
 \\
\hline

\end{tabular}}
\label{tab6:mutability}

\end{table}


The \textbf{ACAC$_2$} model supports features including constraints, pre-, ongoing or post conditions, obligations, and inherits the characteristics from our base ACAC$_0$ model. Constraints supports static and dynamic control of activities as defined.  

\begin{itemize}[leftmargin=*]
  \item \textbf{Static Constraints:} These constraints include the initial pre-defined conditions, the type of activities a device can perform, the attributes a source and target device can possess to start an activity. It also covers activity relationships such as incompatibility \cite{ACAC}. In short, static constraints are predefined scope of the activities that can govern to decide whether an activity can transition to a start state.
  \item \textbf{Dynamic Constraints:} These set of constraints include the dynamic separation of duties from the perspective of the devices or users involved in the invocation of an activity. Dynamic constraints also evaluate the conditions and dependencies among activities to continue or revoke the permission of an activity.  For example, an ongoing condition can be an event (e.g., a wind storm) that triggers an activity (e.g., automatic closing the roof ventilators of home). Continuity and mutability of activity is strongly tied with the dynamic constraints. These constraints are evaluated during run-time.
\end{itemize}
\subsection{ACAC and Zero Trust}
\textbf{Zero trust} is now one of the most used buzzwords and a de-facto requirement in cybersecurity. It is an approach that indicates there is no implicit trust in digital interactions and all of the interactions in cyber systems need to be validated. According to National Institute of Standards and Technology \cite{rose2020zero}, \textbf{“Zero trust architecture (ZTA) is an enterprise’s cybersecurity plan that utilizes zero trust concepts and encompasses component relationships, workflow planning, and access policies”}. Zero Trust architecture is designed considering a number of basic tenets, which are not required to be fully and explicitly implemented for a given strategy. However, with adherence to the core concept of the tenets, a system can be convergent to ZTA. 

\subsubsection{\textbf{Convergence with Zero Trust Architecture}}
Our proposed ACAC model captures the following zero trust tenets \cite{rose2020zero} supporting the next generation requirements of enforcing security posture assuming there is no trust without validation.
\begin{itemize}[leftmargin=*]
  \item \textbf{All data sources and computing services are considered resources:} ACAC objects such as devices, sensors, applications are considered as resources. These resources are producing or collecting the data in the system.
  \item \textbf{All communication is secured regardless of network location:} Network location alone can not make sure a secured access in any system. In ACAC, our main focus is on controlling access regarding the initiation of an activity which may require an operation performed on the corresponding device. This does not depend on the network location rather it depends on dynamic decision components that are discussed in Subsection \ref{modelComponents}.
  \item \textbf{Access to individual enterprise resources is granted on a per-session basis:} Access to an object is granted on per activity-request basis where the execution of an activity can be mapped to a session. A particular session checks specific conditions, changes, other dependent and mutable activities to be in desired states.
  \item \textbf{Access to resources is determined by dynamic policy — including the observable state of client identity, application/service, and the requesting asset—and may include other behavioral and environmental attributes:} ACAC supports dynamic policies that include authorizations (based on source and object attributes and operational right), environmental conditions (comparable to environmental attributes), obligations and states of related dependent activities. These policies determine the access decision on per activity-initiation request.
  \item \textbf{The enterprise collects as much information as possible about the current state of assets, network infrastructure and communications and uses it to improve its security posture:}  ACAC considers different current contextual information (states of devices, system or environmental conditions, source and object attributes) relevant to the decision making which strengthens the security posture of the connected smart systems.
\end{itemize}

\subsubsection{\textbf{Trust Algorithm and Risk Analysis in Future Work of ACAC}}
Policy Engine (PE) is a logical component of zero trust architecture where the ultimate access decision is taken using the policies defined for the cyber system. Trust algorithms are used by the PE that determines the grant, deny or revoke access to the resources. Rose et al. \cite{rose2020zero} talk about trust algorithm variations and show how contextual and score-based algorithms are more dynamic than criteria-based and singular algorithms. In future work, we intend to adapt score-based algorithm by analyzing the risk factors and evaluate a threshold score as confidence level above which score the PE can grant the access to a resource. To make our proposed model more dynamic, past access history to objects will be considered as a factor in activity decision making which is referred in contextual algorithm of ZTA \cite{rose2020zero}.
\vspace{-3mm}
\section{Conclusion and Future Work}
\label{sec:conc}

In IoT-based connected and smart CPS, activities are inseparable part from the process of automation, and require run-time access control to allow of deny a requested activity in a system. 
We envision an \textbf{active} security model focusing on the activity-centric access control for smart collaborative ecosystems. The proposed ACAC model shows how current or preceding activities constraint the initiation of a new activity along with considering few other parameters such as authorizations, obligations and conditions. In this paper, we discuss the distinctions between ACAC and other related access control models. We highlight the factors for why these models are not fit for the systems with wide-range of connected IoT-based smart devices and activities. We present preliminary thoughts on model components, states of an activity and show a hierarchical structure for family of models (ACAC$_0$, ACAC$_1$, ACAC$_2$, and ACAC$_3$) that adds the significant supporting properties gradually to strengthen the model. We briefly discuss how Zero Trust tenets are supported by ACAC model. 

For future work, we will develop formal operational and administrative models, policy language, and enforcement architectures for ACAC, as elaborated Gupta and Sandhu \cite{ACAC}. To support ZTA, we will accommodate the zero trust tenets and trust algorithms by analyzing risks present in the system. We also aim to build a self-adaptive and AI-driven ACAC model that will not need explicit policy definition for each access. AI-driven policy mining from activity logs will make the activity-centric control decision automated for connected smart CPS.

\section*{Acknowledgement}
This research is partially supported by NSF CREST Center Grant HRD-1736209 at UTSA, and by the NSF Grant 2025682 at TTU.

\bibliographystyle{ACM-Reference-Format}
\bibliography{bibliography}

\end{document}